\documentclass{article}
\usepackage{amssymb}
\usepackage{amsmath,bm}
\DeclareMathAlphabet\mathbfcal{OMS}{cmsy}{b}{n}
\DeclareMathAlphabet\mathbfit{OT1}{cmr}{bx}{it}
\numberwithin{equation}{section}

\usepackage{graphicx}

\newtheorem{theorem}{Theorem}[section]

\newtheorem{remark}[theorem]{Remark}

\newcommand{\D}{\mathrm{d}}
\newcommand{\tr}{{\mathrm{tr}}}

\newcommand{\uu}{{\mathbf{u}}}
\newcommand{\ux}{{\mathbf{x}}}

\newcommand{\pp}[2]{ \frac{\partial #1}{\partial #2} }
\begin{document}
\title{On blowups of vorticity for the homogeneous Euler equation}
\author{
B.G.Konopelchenko $^1$ and G.Ortenzi $^{2}$ 
\footnote{Corresponding author. E-mail: giovanni.ortenzi@unimib.it}\\ 
$^1$ {\footnotesize Dipartimento di Matematica e Fisica ``Ennio De Giorgi'', Universit\`{a} del Salento, 73100 Lecce, Italy} \\
 $^2$ {\footnotesize  Dipartimento di Matematica e Applicazioni, 
Universit\`{a} di Milano-Bicocca, via Cozzi 55, 20126 Milano, Italy}\\
$^2${\footnotesize  INFN, Sezione di Milano-Bicocca, Piazza della Scienza 3, 20126 Milano, Italy}
} 
\maketitle
\abstract{
Blowups of vorticity for the three- and two- dimensional homogeneous Euler equations are studied. Two regimes of approaching a blowup points,
respectively, with variable or fixed time are analysed. It is shown that in the $n$-dimensional ($n=2,3$) generic case the blowups of degrees $1,..,n$ at the variable time regime and of  degrees $1/2,..,(n+1)/(n+2)$ at the fixed time regime may exist.  Particular situations when the vorticity blows while the direction of the vorticity vector is concentrated in one or two directions are realisable. 
}
\section{Introduction}
Vorticity and associated phenomena are among the most studied subjects in hydrodynamics (see e.g. \cite{Lamb,L-VI,Bac,Saf,MB} and the papers 
\cite{CLM85,CF93,Tao16,Che91,CC19,KR98,KR00,Kuz02,Kuz03,KM22}). Number of approaches and different techniques have been developed. 
Most of the studies of the blowups of vorticity has been performed for the ideal incompressible fluid. The compressible case is considered as 
the much more  complicated one (see e.g.  \cite{Lamb,L-VI,Bac,Saf,MB,CLM85,CF93,Tao16}).\par

In the papers \cite{Che91} and \cite{Kuz03} it was observed that in the case of compressible fluid the behavior of vorticity for the Euler equation 
is intimately connected with that of homogeneous Euler equation (HEE)
\begin{equation}
\uu_t+\uu \cdot \nabla \uu=0\, .
\label{HEeq}
\end{equation}
without the constraint $\nabla \cdot \uu=0$. In the papers \cite{Che91,CC19} an explicit integral-type formula for the vorticity 
$\boldsymbol{\omega}=\nabla \times \uu$
for the equation (\ref{HEeq}) has been presented. Another type of formula for the vorticity has been found in \cite{KR98,KR00}. 
The blowup of vorticity as $t \to t_c>0$ has been analysed in \cite{KR00,Kuz03} (see also \cite{CC19} and  \cite{KM22}). \par

Homogeneous Euler equation (\ref{HEeq}) is the most simplified version of the basic equations of the hydrodynamics when  one can neglect 
all effects of pressure, 
viscosity etc.. Nevertheless  it has number of applications in physics and represent itself an excellent touchstone for an analysis of blowups of vorticity. \par

In this paper we  present some results concerning the blowups of vorticity for the three- and two-dimensional homogeneous 
Euler equation (\ref{HEeq}).
Our analysis is based in part  on the previous study of the structure and hierarchies of blowups of derivatives for the $n$-dimensional HEE \cite{KO22,KOpre}. \par

We consider the behavior of vorticity in two different regimes of approaching the blowup points at the blowup hypersurface. The first regime 
is to approach
such a point along the $t$ axis, i.e. $t \to t_b$ while the coordinates $\uu$ in the hodograph space remain fixed. It is shown that, in the generic case, 
i.e. for generic initial data for the 3D HEE (\ref{HEeq}), the vorticity in this regime may have singularities of three different degrees
\begin{equation}
\omega_i \sim (t-t_b)^{-m}\, , \qquad t \to t_b, \qquad m=1,2,3\, .
\label{time-ufix-bu}
\end{equation}
Such blowups occur on the intersection of $m$ branches of the blowup hypersurface $\Gamma$. The existence of blowups of type (\ref{time-ufix-bu})
with $m=1,2$ has been observed earlier in \cite{KR00}. 


In the second regime the time $t_b$ is fixed while the coordinates $\uu$ are variyng. In this regime  of approaching the blowup point for 
3D HEE (\ref{HEeq}) generically it may exists 4 levels of blowups of the vorticity 
$\boldsymbol{\omega}$ with the behavior
\begin{equation}
\omega_i \sim \epsilon^{-\frac{m}{m+1}}\, , \qquad m=1,2,3,4\, ,
\label{vort-bu}
\end{equation}
where $\epsilon \sim |\delta \ux| \to 0$. Blowups (\ref{vort-bu}) occur on the subspaces $\Gamma_m$ of the blowup hypersurface $\Gamma$ and 
dim$\Gamma_m=4-m$, $m=1,2,3,4$. \par

It may happens also that the components of the vorticity $\boldsymbol{\omega}$ behave differently on certain subspaces of $\Gamma$. 
In particular, at the first level $m=1$ there may exist one-dimensional subspace ${\Gamma}_1$ at which 
the component $\omega_3$ blows as $\epsilon^{-1/2}$ when $\epsilon \to 0$ while the components
 $\omega_1$ and $\omega_2$ remain bounded. 
In such a case the direction of the vorticity $\boldsymbol{\hat{\omega}}$ is a unit vector oriented along one axes, namely
\begin{equation}
\boldsymbol{\hat{\omega}} = \big( 0,0,1\big)\, . 
\end{equation}
The calculations are performed both in the special 
coordinates introduced in \cite{KOpre} as well as in cartesian coordinates $\ux$ and $\uu$.\par

For the 2D HEE (\ref{HEeq}) the vorticity blows-up as in (\ref{time-ufix-bu}) and (\ref{vort-bu}) with $m$ taking the values $m=1,2$ for 
 (\ref{time-ufix-bu}) and $m=1,2,3$ for (\ref{vort-bu}), respectively. Three particular solutions of the 2D HEE with different blowup
 behavior are considered.

It is noted that we analyze the behavior of vorticity at certain points on the blowup hypersurface $\Gamma$ and at the time $t_b$ which can be
negative or positive. The realisability of blowups of different orders at positive time remains an open problem. \par

Similar results for the $n$-dimensional homogeneous Euler equation are briefly discussed too.

The paper is organized as follows. Section \ref{sec-der-bu} contains a brief exposition of the results of the paper \cite{KOpre} for the 3D HEE. 
Blowups of vorticity in the first regime $t \to t_b$ are analysed in section \ref{sec-vort-bu-var-t}.
Blowups of vorticity for the 3D HEE in the regime with fixed $t$ are studied in sections \ref{sec-vort-bu} and \ref{sec-up-bu}. 
Similar results for the 2D HEE are presented in section \ref{sec-vort2D}.  
Three particular solutions of the 2D HEE with different blowup behavior are described in detail in Section \ref{sec-vortexe2D}.
The $n$-dimensional $n \geq 4$ case is discussed in Section \ref{sec-n-dim}.
Conclusion \ref{sec-Con} contains some indications on possible future developments.
\section{Blowups of derivatives}
\label{sec-der-bu}
Here for conveniency we report some results concerning the blowup of derivatives for the 3-dimensional homogeneous Euler equation obtained in the paper 
\cite{KOpre}. We also slightly change the notations in order to make the corresponding formulas  more convenient for the further calculations.\par

The starting point of the analysis are the hodograph equations \cite{Zel70,Che91,Fai93,KO22} 
\begin{equation}
x_i=u_i t +{f_i}(\uu)  \, , \qquad i=1,2,3
\label{hodogen}
\end{equation}
where $f_i(\uu)$ are arbitrary functions locally inverse to the initial data $u_i(t=0,\ux)$. 
Any solution $\uu(\ux,t)$ of the system (\ref{hodogen}) is a solution of the 3D system
HEE (\ref{HEeq}).\par

The matrix $M$ with the elements
\begin{equation}
M_{ij}=t \delta_{ij}+ \pp{f_i}{u_j} \, , \qquad i,j=1,2,3\, ,
\end{equation}
plays a central role in the analysis of blowups  of derivatives and possible gradient catastrophes.
In particular,
\begin{equation}
 \pp{u_j}{x_k}=(M^{-1})_{jk}\, , \qquad i,k=1,2,3
 \label{derivel}
\end{equation}
The blowups occur on the $3D$ hypersurface $\Gamma$  defined by the equation
\begin{equation}
\det M(t; \uu) = t^3 + a_2(\uu)t^2+ a_1(\uu) t +a_0(\uu)=0\, ,
\label{BU-cond}
\end{equation}
where  $a_2(\uu)$, $a_1(\uu)$ are certain functions of $\uu$ and 
$a_0(\uu)=\det(M(t=0,\uu)) \neq 0$ for generic initial data.\par

The blowup hypersurface $\Gamma$ is the union of the branches $t_\alpha=\phi_\alpha(\uu)$ corresponding to real roots of the cubic equation
(\ref{BU-cond}). In the three dimensional case the number of branches can be one or three \cite{KOpre}.\par 

In the generic case the rank $r$ of the matrix $M$ may assume two values $r=2$ and $r=1$. Equivalently it means that there exists $3-r$ vectors 
$\mathbf{R}^{(\alpha)}(\uu_b)$ and $\mathbf{L}^{(\alpha)}(\uu_b)$\,  , $\alpha=1,3-r$ such that ($\uu_b \in \Gamma$)
\begin{equation}
\begin{split}
\sum_{j=1}^3 M_{ij} R^{(\alpha)}_j &=0\, ,\qquad i=1,2,3 \, , \quad \alpha=1,3-r\, , \\
\sum_{i=1}^3 L^{(\alpha)}_i  M_{ij} &=0\, ,\qquad  j=1,2,3 \, , \quad \alpha=1,3-r \, .
\end{split}
\label{RLeigen}
\end{equation}
The existence of such vectors suggests the introduction of new dependent and independent variables $v_1,v_2,v_3$ and $y_1,y_2,y_3$ 
defined by the relations
\cite{KOpre}
\begin{equation}
\begin{split}
\delta \uu \equiv &  \sum_{\alpha=1}^{3-r} \mathbf{R}^{(\alpha)} \delta v_\alpha +  \sum_{\beta=1}^{r} \mathbf{\tilde{R}}^{(\beta)} \delta v_{\beta+3-r} 
\equiv \sum_{\alpha=1}^3 \mathbfcal{R}^{(\alpha)}  \delta v_\alpha\, ,  \\
\delta \ux \equiv &  \sum_{\alpha=1}^{3-r} \mathbf{P}^{(\alpha)} \delta y_\alpha +  \sum_{\beta=1}^{r} \mathbf{\tilde{P}}^{(\beta)} \delta y_{\beta+3-r} 
\equiv \sum_{\alpha=1}^3 \mathbfcal{P}^{(\alpha)}  \delta y_\alpha  \, ,
\end{split}
\label{bu-var}
\end{equation}
where the vectors $\mathbf{\tilde{R}}^{(\beta)}$ are $r$ vectors complementary to the set of $3-r$ vectors $\mathbf{R}^{(\alpha)}$ and
vectors $P^{(\alpha)}$, $\tilde{P}^{(\beta)}$ are defined by the relation 
\begin{equation}
 \sum_{\alpha=1}^{3-r} {P}^{(\alpha)}_i {L}^{(\alpha)}_j+ \sum_{\beta=1}^{r} {\tilde{P}}^{(\beta)}_i {\tilde{L}}^{(\beta)}_j 
 =\delta_{ij}\, , \qquad i,j=1,2,3
 \label{def-inv-left-eigen}
\end{equation}
where $\mathbf{\tilde{L}}^{(\beta)}$ are $r$ vectors complementary to the set of $3-r$ vectors $\mathbf{L}^{(\alpha)}$.
One also has 
\begin{equation}
\delta y_\beta=\mathbfcal{L}^{(\beta)} \cdot \delta \ux =\sum_{i,j=1}^3\mathcal{L}^{(\beta)}_i M_{ij}(\uu_b) \mathcal{R}^{(\alpha)}_j \delta v_\alpha+ O(|\delta v|^2)\, ,
\label{map-adapt-cart}
\end{equation}
where $\sum_{\alpha=1}^{3} \mathcal{P}^{(\alpha)}_i \mathcal{L}^{(\alpha)}_j =\delta_{ij}$. \par

The use of variational consequences of the hodograph equations (\ref{hodogen}) shows that derivatives $\pp{v_\alpha}{y_\beta}(\uu_b)$ behave differently
in different subsectors of the independent and dependent variables \cite{KO22,KOpre}. For instance, for $r=2$, on the first level of blows-ups, 
the derivatives 
\begin{equation}
\pp{v_1}{y_1}\, , \pp{v_1}{y_2}\, ,  \pp{v_1}{y_3}\, , \pp{v_2}{y_1}\, ,\pp{v_3}{y_1} 
\label{siBU}
\end{equation}
explode on the hypersurface $\Gamma$ (\ref{BU-cond}) while the derivatives
 \begin{equation}
\pp{v_2}{y_2}\, , \pp{v_2}{y_3}\, ,   \pp{v_3}{y_2}\, ,\pp{v_3}{y_3}
\label{noBU}
\end{equation}
remain bounded. These blowups may happen both at positive and negative time.

It is noted that all vectors given above  and the behavior of derivatives $\pp{v_\alpha}{y_\beta}$ vary with the variation of the point $\uu_b$ 
belonging to the hypersurface $\Gamma$ (\ref{BU-cond}). \par

On the first level of blows-ups the derivatives explode as $\epsilon^{-1/2}$,  $\epsilon \sim |\delta y| \to 0$ and the behavior of derivatives at fixed time 
$t_b$ presented in (\ref{siBU}) and (\ref{noBU}) can be resumed in the formula
\begin{equation}
\delta v_\alpha \sim \sum_{j=1}^3 C_{\alpha \beta} \delta y_\beta 
\, , \qquad i=1,2,3\, , \\
\label{bu-beha-comp}
\end{equation}
where
\begin{equation}
C=
\begin{pmatrix}
 \epsilon^{-1/2}  \nu_{11} & \epsilon^{-1/2} \nu_{12} & \epsilon^{-1/2}  \nu_{13} \\
 \epsilon^{-1/2}  \nu_{12} &  \nu_{22}  &  \nu_{23} \\
 \epsilon^{-1/2}  \nu_{13} &  \nu _{32} &  \nu_{33} 
\end{pmatrix}
\label{bu-beha}
\end{equation}
and $\nu_{ij}$, $i,j=1,2,3$  are connected with the values of $\pp{f_i}{u_j}(\uu_b)$ and $\frac{\partial^2 f_i}{\partial u_j \partial u_k}(\uu_b)$ evaluated at the point 
$\uu_b \in \Gamma_1$ (see \cite{KOpre}). 

We emphasize that the formulae (\ref{bu-beha}) represent the relations between the infinitesimal variations of the variables $y_i$ and $v_i$ around a point
$\uu_b \in  \Gamma$ at fixed time $t_b$. Blowup time $t_b$ can be positive or negative. Blowup at $t_b>0$ is refereed as gradient catastrophe. 
In this paper, as in \cite{KOpre}, we will not discuss conditions which guarantee that $t_b>0$. \par

It is also noted the domain $\mathcal{D}_\uu$ of variations of $\uu$ constructed {\it via} equation (\ref{hodogen}) and, consequently, the domain 
of variations of variables $\uu$ parameterizing the blowup hypersurface $\Gamma$ (\ref{BU-cond}),
\begin{equation}
\mathcal{D}_\uu \equiv \{ \uu : \det M(t,\uu)=0\}\, ,
\end{equation}
 coincides with the domain $\mathcal{D}_{\uu_0}$ 
of variations of the initial values $\uu_0$, since $\uu(\ux,t)=\uu_0(\ux-\uu t)$.


\section{Blowup of vorticity}
\label{sec-vort-bu-var-t}
The formula (\ref{derivel}) provide us with the explicit and useful expression for the vorticity vector in the original Cartesian coordinates in terms of
the components $u_i$, $i=1,2,3$ of the velocity. Namely,
\begin{equation}
\begin{split}
\omega_i&= \sum_{j,k=1}^3 \epsilon_{ijk} \pp{u_k}{x_j}= \sum_{j,k=1}^3 \epsilon_{ijk} (M^{-1})_{kj}= 
\frac{1}{\det(M(t,\uu))}\sum_{j,k=1}^3 \epsilon_{ijk} \widetilde{M}_{kj}(t,\uu)
\, , \qquad     i=1,2,3
\end{split}
\label{vortivel}
\end{equation}
where $\widetilde{M}$ is the adjugate matrix. \par

We consider first the case $\mathrm{rank} (M(t_b,\uu_b))=2$. Let us fix the point $\uu_{b}$ on the blow-up hypersurface $\Gamma$ (\ref{BU-cond}) and take the corresponding real $t_{b}$, i.e. the real root
of the cubic equation(\ref{BU-cond}) which always exists for the 3D HEE \cite{KO22}. The formula (\ref{vortivel}) implies that 
(\cite{KO22})
\begin{equation}
\omega_i(t=t_b+\varepsilon)\Big\vert_\Gamma
= \frac{\sum_{j,k=1}^3  \epsilon_{ijk} \left( \widetilde{M}_{kj} (t_b,\uu_b) + \varepsilon  \widetilde{M}'_{kj}(t_b,\uu_b) + O(\varepsilon^2) \right) }
{\varepsilon D_1(t_b,\uu_b)+\varepsilon^2 D_2(t_b,\uu_b)+ \varepsilon^3}\, , \qquad \varepsilon \equiv t-t_b\to 0
\label{vortivel-dev}
\end{equation}
where  
\begin{equation}
\begin{split}
D_1&\equiv \pp{\det(M(t,\uu))}{t}\Big\vert_{t_b,\uu_b}= 3t_b^2+2a_2(\uu_b)t_b+a_1(\uu_b)\, ,\\
D_2&\equiv \frac {\partial ^2 \det(M(t,\uu))}{\partial t^2 }\Big\vert_{t_b,\uu_b}=3t_b+a_2(\uu_b)\, , \\
\end{split}
\label{coeff-vortiveldev}
\end{equation}
and $\widetilde{M}'_{kj}(t_b,\uu_b) \equiv \frac{\D \widetilde{M_{kj}}(t,\uu)}{\D t}\Big{\vert}_{t_b,\uu_b}$.
Generically for $r=2$ $\widetilde{M}_{jk}(t_b,\uu_b) \neq 0$ and $D_1(t_b,\uu_b) \neq 0$. 
Hence, in the generic case, in the first regime the vorticity  blows up on the full hypersurface $\Gamma$  as
\begin{equation}
\omega_i(t,\uu_b) \sim \sigma_i \varepsilon^{-1} 
\equiv \sigma_i (t-t_b)^{-1}\, , \qquad t \to t_b\, , \quad i=1,2,3
\label{vort-t-1}
\end{equation}
where $\sigma_i \equiv {\sum_{j,k=1}^3\epsilon_{ijk} \widetilde{M}_{kj}(t_b, \uu_b)}/{D_1(t_b,\uu_b)}$ for $i=1,2,3$.\par

Existence of the higher order singularities is correlated with the structure of the blowup hypersurface $\Gamma$. If it has a single branch
(single real root of the equation (\ref{BU-cond})) then $M'(t_b,\uu_b)$ cannot be zero. Hence, due to 
(\ref{vortivel-dev}) and (\ref{coeff-vortiveldev}) in this case only the blowup of type (\ref{vort-t-1}) occurs.\par

Situation is different when $\Gamma$ has three real branches, i.e. all roots of the equation (\ref{BU-cond})  are real. 
In this case one has the formulae (\ref{vortivel-dev}) and (\ref{coeff-vortiveldev}) and three different values of ${t_b}_\alpha$, $\alpha=1,2,3$ for the 
same value  $\uu_b$. Moreover, the condition
\begin{equation}
\pp{\det(M(t_b,\uu_b))}{t}=0\, ,
\end{equation}
i.e. the condition that $\det(M(t_b,\uu_b))$ has a double zero at $t_b$, is now admissible.

 Let the condition 
 \begin{equation}
 D_1(t_b,\uu_b)=3 t_b^2 +2a_2(\uu_b)t_b+a_1(\uu_b)=0
 \label{cond-vort-2}
 \end{equation}
 be satisfied at one branch. It defines the two-dimensional submanifold $\mathcal{D}_\uu^{(2)}$ at $\mathcal{D}_\uu$. 
 At fixed $\uu_b \in \mathcal{D}_\uu^{(2)}$  and at the corresponding ${t_b}_\alpha$ 
the vorticity blows-up as
\begin{equation}
\omega_i(t,\uu_b) \sim \varepsilon^{-2}\equiv(t-t_b)^{-2}\, , \qquad t \to t_b\, .
\label{vort-t-2}
\end{equation}

Moreover, the condition (\ref{cond-vort-2}) (cf. (\ref{coeff-vortiveldev})) means that the root ${t_b}_\alpha$ is a double root, 
i.e. coincides with another root  ${t_b}_\beta$.
So, the branches $\alpha$ and $\beta$ of the blowup hypersurface $\Gamma$ intersect along the two-dimensional surface $\Gamma_2$  
corresponding to values of $\uu_b \in \mathcal{D}_\uu^{(2)}$ and on $\Gamma_2$ the vorticity blows up as in (\ref{vort-t-2}). \par

Hence, in the particular case (\ref{cond-vort-2}) the vorticity $\boldsymbol{\omega}$ blows up ad $(t-t_b)^{-2}$ on the intersection of two branches
of $\Gamma$  and blows up  as $(t-t_b)^{-1}$ on the third branch.\par


Finally if, in addition to (\ref{cond-vort-2}),  the condition
 \begin{equation}
 D_2(t_b,\uu_b)=3 t_b +a_2(\uu_b)=0\, , 
 \label{cond-vort-3}
 \end{equation}
 is satisfied, but
 $\sum_{j,k=1}^3  \epsilon_{ijk}  \widetilde{M}_{kj} (t_b,\uu_b)   \neq 0 $, with $i=1,2,3$ 
 then the vorticity $\boldsymbol{\omega}$ blows up as
\begin{equation}
\omega_i(t,\uu_b) \sim \varepsilon^{-3}\equiv(t-t_b)^{-3}\, , \qquad t \to t_b\, .
\label{vort-t-3}
\end{equation}

The situation (\ref{vort-t-3}) happens on the curve $\Gamma_3$ in $\mathcal{D}_\uu$ defined by the conditions (\ref{cond-vort-2}) and (\ref{cond-vort-3}). 
Since such conditions means that the root ${t_b}_\alpha$ is a triple root, the behavior (\ref{vort-t-3}) occurs at the intersection of all three branches of the blowup  surface $\Gamma$.\par

Other possible  situations, for instance, the condition $D_1({t_b}_\alpha,\uu_b)=0$ for all $\alpha=1,2,3$ are equivalent to 
(\ref{cond-vort-3}) and  (\ref{vort-t-3}). \par

The existence of the blowups of the types (\ref{vort-t-1}), (\ref{vort-t-2}),  and (\ref{vort-t-3}) becomes rather obvious 
if one rewrites the formula (\ref{BU-cond}) as 
\begin{equation}
\det \left( M(t,\uu) \right)= (t-{t_b}_1)(t-{t_b}_2)(t-{t_b}_3)\, .
\label{BU-cond-zero}
\end{equation}

It is noted that one can treat the conditions (\ref{cond-vort-2}) and (\ref{cond-vort-3}) in a different manner, namely, to consider them as the equations 
for the functions $f_1(\uu),f_2(\uu),f_3(\uu)$. Within such a viewpoint, equation (\ref{cond-vort-2}) defines those functions $f_i(\uu)$, $i=1,2,3$ for which 
two branches of the hypersurface $\Gamma$ identically coincide. All three branches of $\Gamma$ coincide in the particular case of initial data 
such that the functions $f_i(\uu)$, $i=1,2,3$ are solutions of the pair of equations (\ref{cond-vort-2}) and (\ref{cond-vort-3}).\par

The formulae (\ref{vort-t-1}) and (\ref{vort-t-2}) reproduce the results previously obtained in \cite{KR00} with the use of the Lagrangian analogue 
of the formula (\ref{vortivel}). The behavior of type (\ref{vort-t-3}) was not present in  \cite{KR00} due to the particular 
geometry of the vortex lines considered there. \par

An analysis of the behavior of vorticity and its integral characteristics has been performed also in \cite{CC19}  with the use of an 
explicit integral representation of the Lagrangian type derived in \cite{Che91}.\par



The components $\omega_i$ behave according to (\ref{vort-t-1}), (\ref{vort-t-2}),  and (\ref{vort-t-3}) in the general case when all 
$\sigma_i \neq 0$. In this case the direction of the vorticity vector 
(see e.g.\cite{CF93})
\begin{equation}
\boldsymbol{\hat{\omega}}\equiv \frac{\boldsymbol{\omega}}{|\boldsymbol{\omega}|}  
\label{vort-bu-dir}
\end{equation}
is regular with components 
\begin{equation}
\boldsymbol{\hat{\omega}}= \frac{1}{|\mathbf{\sigma}|} (\sigma_1,\sigma_2,\sigma_3)\, , \qquad  |\mathbf{\sigma}|^2=\sigma_1^2+\sigma_2^2+\sigma_3^2\, .
\end{equation}

Let us assume now that one of $\sigma_i$ vanishes, e.g. $\sigma_3$, i.e.
\begin{equation}
\sum_{j,k=1}^3\epsilon_{3jk} \widetilde{M}_{kj}(t_b, \uu_b)=0\, .
\end{equation}
This condition defines the two-dimensional subspace $\mathcal{D}_2 \subset \mathcal{D}_\uu$ in the hodograph space. At the points 
$\uu \in \mathcal{D}_2 $  one has $\sigma_3 =0$ and, hence instead of (\ref{vort-t-1}) the vorticity vector direction blows up as  as
\begin{equation}
\omega_1 \sim \sigma_1 (t-t_b)^{-m}\, , \qquad
\omega_2 \sim \sigma_2 (t-t_b)^{-m}\, , \qquad
\omega_3 \sim \sigma'_3 (t-t_b)^{-m+1} \, , \qquad m=1,2,3\, ,
\end{equation}
where $\sigma_i' \equiv \sum_{j,k=1}^3\epsilon_{ijk} \widetilde{M}'_{kj}(t_b, \uu_b)$ for $i=1,2,3$. Consequently, the vector 
$\boldsymbol{\hat{\omega}}$ is of the form
\begin{equation}
\boldsymbol{\hat{\omega}}=\frac{1}{|\boldsymbol{\sigma}|} (\sigma_1,\sigma_2,0)\, .
\end{equation}
Generically, for $m=1$ such situation may occur on the two-dimensional subsurface of the blow-up hypersurface $\Gamma$. For $m=2$ it may happens
along the curve belonging to the two-dimensioanl intersection of two-branches of $\Gamma$. For $m=3$ it may occur at the point belonging to 
the curve of intersection of the three branches of $\Gamma$.\par

In the very particular case of two vanishing components of $\sigma_i$, e.g. $\sigma_1=\sigma_2=0$, one has
\begin{equation}
\omega_1 \sim \sigma'_1 (t-t_b)^{-m+1}\, , \qquad
\omega_2 \sim \sigma'_2 (t-t_b)^{-m+1}\, , \qquad
\omega_2 \sim \sigma_3 (t-t_b)^{-m} \, , \qquad m=1,2,3\, ,
\end{equation}
and 
\begin{equation}
\boldsymbol{\hat{\omega}}=(0,0,1)\, .
\end{equation}
Generically such behavior may exists only for $m=1,2$. For $m=1$ it may happens along a curve on $\Gamma$ while for $m=2$
it may occur at the point belonging to the intersection of two branches of $\Gamma$. \par

The behaviour of vorticity described above corresponds to the case of rank $r =2$ for the matrix $M$ evaluated on the blowup hypersurface   
 $\Gamma$. It occurs on the whole blowup hypersurface \cite{KOpre}. 
    In contrast, the matrix $M(t_b, \uu_b)$ may have rank 1 only on a set of points  $\Gamma_0$ on $\Gamma$ \cite{KOpre}. Moreover for $r=1$ 
    the adjugate matrix $ \widetilde{M}$ vanishes identically:  
    \begin{equation}
    \widetilde{M}_{ij}  \Big{\vert}_{\Gamma_0}= 0, \qquad i , j=1,2,3.  
    \end{equation}
    On the other hand generically  $\widetilde{M}_{ij}'  \big{\vert}_{\Gamma_0}$ are different from zero.  
    So, in such a situation the components of vorticity remain bounded when  $t$ is approaching $t_b$ which correspond to  a point $\uu_b$ 
    belonging to $\Gamma_0$.


\section{Blowups of vorticity at fixed time}
\label{sec-vort-bu}
The formulae (\ref{vort-t-1}), (\ref{vort-t-2}), and (\ref{vort-t-3}) describe the behavior of the vorticity in the situation when time $t$ approach the blowup 
time $t_b$ along the $t$ axis with fixed coordinate $\uu_b$.\par

The approach presented in \cite{KOpre} and briefly reproduced in the section \ref{sec-der-bu} looks more appropriate for the analysis 
of the blowups of vorticity
in the regime when time $t$ is fixed while the coordinates $\uu$ are subject to variations.


The formulas presented in the section \ref{sec-der-bu} (see also \cite{KOpre}) indicate that non-cartesian coordinates 
$y_i$ and $v_i$, $i=1,2,3$ are rather 
convenient for the analysis of blowups of the derivatives.  In order to use such coordinates for the analysis of blowups of vorticity, one has to 
consider its coordinate-independent definition as the differential  two-form (see e.g. \cite{AK,Tao16}) 
\begin{equation}
\omega= \D \theta=\D \uu \wedge \D \ux \, . 
\label{vort-2form}
\end{equation}
where $\theta=\uu \cdot \D \ux$.

We will use such definition  in the form
\begin{equation}
\omega(\uu_b)=\delta \uu \wedge \delta \ux\, 
\end{equation}
to study the behavior of vorticity at the point $\uu_b$ of the blowup hypersurface $\Gamma$. 

Using the formulae (\ref{bu-var}), one gets
\begin{equation}
{\omega}(\uu_b) \equiv \sum_{\alpha,\beta=1}^3 q_{\alpha \beta} \delta {v}_\alpha \wedge \delta{y}_\beta\, ,
\end{equation}
where
\begin{equation}
q_{\alpha \beta} \equiv \mathbfcal{R}^{(\alpha)} \cdot \mathbfcal{P}^{(\beta)} \, , \qquad \alpha,\beta=1,2,3\, .
\label{vort-bu-coeff-velpos}
\end{equation}
Then, due to the relation (\ref{bu-beha-comp}), at the blowup point $\uu_b$ one obtains
\begin{equation}
\omega(\uu_b) = \sum_{\alpha,\beta=1}^3 \omega_{\alpha \beta}(\uu_b)  \delta y_\alpha \wedge \delta y_\beta
\label{vort-bu-2f-adapcoord}
\end{equation}
where
\begin{equation}
 \omega_{\alpha \beta}(\uu_b) \equiv \frac{1}{2} \sum_{\gamma=1}^3 (C_{\gamma \alpha} q_{\gamma \beta} - C_{\gamma \beta} q_{\gamma \alpha})\, , \qquad
 \alpha,\beta=1,2,3\, .
 \label{coeff-vort-bu-2f-adapcoord}
\end{equation}
The components of the vorticity vector $\boldsymbol{\omega}$ in these coordinates are defined as usual as
\begin{equation}
\omega_\alpha =\sum_{\beta,\gamma=1}^3 \epsilon_{\alpha \beta \gamma} \omega_{\beta \gamma} \, , \qquad \beta,\gamma=1,2,3\, .  
\end{equation}
At the first level of blowup and rank $r=2$ the matrix $C$ is of the form (\ref{bu-beha}). Consequently, the element of $\omega_{\alpha \beta}$, 
written in terms of 
the vorticity  components $\omega_i$, behave as
\begin{equation}
\omega = \frac{1}{2}
\begin{pmatrix}
0 &  \omega_3 & -\omega_2 \\
- \omega_3&0& \omega_1\\
 \omega_2& -\omega_1&0
\end{pmatrix}
\end{equation}
where
\begin{equation}
\begin{split}
\omega_1 & =  \epsilon^{-1/2} S_1(\uu_b) +T_1(\uu_b)\, ,\\
\omega_2 & =  \epsilon^{-1/2} S_2(\uu_b) +T_2(\uu_b)\, , \\
\omega_3 & =  \epsilon^{-1/2} S_3(\uu_b) +T_3(\uu_b) \, , \\
\end{split}
\label{vort-sing}
\end{equation}
as $\epsilon \to 0$ and
\begin{equation}
\begin{split}
S_1 &  =    (\nu _{1 2} q_{1 3}-\nu _{1 3} q_{1 2}) \, , \qquad  T_1= (\nu _{2 2} q_{2 3}-\nu _{2 3} q_{2 2}+\nu _{3 2} q_{3 3}-\nu _{3 3} q_{3 2}) \, , \\
S_2 & =    (\nu _{1 1} \left(-q_{1 3}\right)+\nu _{1 3} q_{1 1}-\nu _{2 1} q_{2 3}-\nu _{3 1} q_{3 3}) \, , \qquad  T_2=(\nu _{2 3} q_{2 1}+\nu _{3 3} q_{3 1}) \, ,\\
S_3 & =    (\nu _{1 1} q_{1 2}-\nu _{1 2} q_{1 1}+\nu _{2 1} q_{2 2}+\nu _{3 1} q_{3 2}) \, , \qquad  T_3 =(-\nu _{2 2} q_{2 1}-\nu _{3 2} q_{3 1}) \, .\\
\end{split}
\label{vort-sing-comp}
\end{equation}
So, generically,  i.e. when all $S_\alpha \neq 0$, the vorticity $\boldsymbol{\omega}$ blows-up as $\epsilon^{-1/2}$, $\epsilon \to 0$ at the point $\uu_b$ 
of the three-dimensional blowup hypersurface $\Gamma$. 
In this case the direction of the vorticity vector  $\boldsymbol{\hat{\omega}}$
is regular with the components
\begin{equation}
\boldsymbol{\hat{\omega}} = \frac{1}{|\mathbf{S}|} \Big( S_1(\uu_b),S_2(\uu_b),S_3(\uu_b) \Big)\, .
\end{equation}

However, particular situations are also admissible. Indeed, if there exist  a point $\uu_b \in \Gamma$ such 
that $S_3(\uu_b)=0$ then at this point the components $\omega_1$ and $\omega_2$ of the vorticity blowup while the component $\omega_3$ remain finite. 
The condition $S_3(\uu_b)=0$ has co-dimension one. So, such situation is realisable, in principle, on the two-dimensional sub-surface  of the blowup hypersurface $\Gamma$ and $\boldsymbol{\hat{\omega}}$ is of the form
\begin{equation}
\boldsymbol{\hat{\omega}} = \frac{1}{|\mathbf{S}|} \Big( S_1(\uu_b),S_2(\uu_b),0 \Big)\, .
\end{equation}
\par

Further, there may exist the points belonging to a certain curve on $\Gamma$ at which 
\begin{equation}
S_1(\uu_b)=S_2(\uu_b)=0\, .
\label{deg-gen-vort}
\end{equation}
At these points the components $\omega_1$ and $\omega_2$ remain bounded and only one component $\omega_3$ of the vorticity blows up.
Hence, the vorticity direction vector (\ref{vort-bu-dir}) assumes a particular form 
\begin{equation}
\boldsymbol{\hat{\omega}} = \big( 0,0,1 \big)\, .
\end{equation}

Such a situation when the vorticity vector $\boldsymbol{\omega}$ becomes very large in modulus, but concentrated in one direction looks rather special and of interest. \par

It may even happens at a certain point $\uu_b \in \Gamma$ that
\begin{equation}
S_1(\uu_b)=S_2(\uu_b)=S_3(\uu_b)=0\, .
\end{equation}
In such a case the vorticity $\boldsymbol{\omega}$ remains bounded in the point of the first level blowups of derivatives. \par

Finally, in order to analyse the blowup of vorticity in the Cartesian coordinates it is sufficient to perform the change of coordinates 
$\mathbf{y} \to \mathbf{x}$  in the r.h.s. of (\ref{vort-bu-2f-adapcoord}). 
Performing the transformation (\ref{map-adapt-cart}) in (\ref{vort-bu-2f-adapcoord}), one obtains
\begin{equation}
\omega(\uu_b) = \sum_{i,j=1}^3 \omega_{ij}(\uu_b)  \delta x_i \wedge \delta x_j\, .
\label{vort-bu-2f-cartcoord}
\end{equation}
As a result, the components $\omega_i = \sum_{j,k=1}^3\epsilon_{ijk}\pp{u_k}{x_j}$ of the vorticity vector 
$\boldsymbol{\omega}=\nabla \times \uu$ blows-up on the whole hypersurface $\Gamma$, namely
\begin{equation}
\omega_i= \epsilon^{-1/2} \tilde{S}_i(\uu_b) +\tilde{T}_i(\uu_b) \, , \qquad i=1,2,3\, , \qquad \epsilon \to 0\,  
\label{vort-cart-comp}
\end{equation}
where $\tilde{S}_i$ and $\tilde{T}_i$ are bounded functions obtained by a change of variables from (\ref{vort-sing-comp}).\par

The same result can be obtained directly, using the formulae (\ref{bu-var}), (\ref{bu-beha-comp}) and (\ref{map-adapt-cart}). Namely, one gets 
\begin{equation}
\pp{u_l}{x_k}=\sum_{\beta=1}^3 \pp{u_l}{y_\beta}\pp{y_\beta}{x_k}= \sum_{\alpha,\beta=1}^3 \mathcal{R}^{(\alpha)}_l \pp{v_\alpha}{y_\beta} \mathcal{L}^{(\beta)}_k 
= \sum_{\alpha,\beta=1}^3 \mathcal{R}^{(\alpha)}_l C_{\alpha \beta} \mathcal{L}^{(\beta)}_k \, , \qquad l,k=1,2,3\, ,
\end{equation}
and, then one obtains the formula (\ref{vort-cart-comp}).\par

Again, it may happens that along certain curves $\Gamma_1$ belonging to $\Gamma$, one has
\begin{equation}
\tilde{S}_1(\uu_b)=\tilde{S}_2(\uu_b)=0\, .
\end{equation}
At the points on this curve, the components $\omega_1$ and $\omega_2$ remain bounded while the component $\omega_3 \to \infty$ and
$\hat{\boldsymbol{\omega}}=(0,0,1)$. Such a situation, when
the vorticity vector $\boldsymbol{\omega}$ becomes very large in modulus but concentrated in one direction, resembles somehow certain 
well-known physical phenomena. 

\section{Blowups at rank 1 and higher levels}
\label{sec-up-bu}
In the case of rank $r=1$, which occurs at a set of points $\uu_b \in \Gamma$ the matrix $C$ is of the form (cf. \cite{KOpre})
\begin{equation}
C=\begin{pmatrix}
 \epsilon^{-1/2}  \mu_{11} & \epsilon^{-1/2} \mu_{12} & \epsilon^{-1/2}  \mu_{13} \\
 \epsilon^{-1/2}  \mu_{12} &  \epsilon^{-1/2} \mu_{22}  & \epsilon^{-1/2} \mu_{23} \\
 \epsilon^{-1/2}  \mu_{13} & \epsilon^{-1/2}  \mu _{32} &  \mu_{33} 
\end{pmatrix}\, .
\end{equation}
The components of the vorticity vector $\boldsymbol{\omega}$ again are of the form (\ref{vort-sing-comp}) or (\ref{vort-cart-comp}). \par

However in this case one 
cannot impose any constraint of the type $S_3=0$ or (\ref{deg-gen-vort}), if one considers the situation with generic function $f_i(\uu)$ of initial data. Such
constraints may be admissible for particular special initial data.\par
Blowups of second, third and fourth level for $r=2$ occur on certain subspaces of the three-dimensional blowup hypersurface $\Gamma$ \cite{KOpre}.\par

One of the subsections of the second level of blowups (in the rank $2$ case)  is characterized by the following behavior of derivatives \cite{KOpre}
\begin{equation}
\begin{split}
\pp{v_1}{y_1} \sim \epsilon^{-2/3}\, , \qquad \pp{v_1}{y_2} , \pp{v_2}{y_1} ,\pp{v_1}{y_3} , \pp{v_3}{y_1} \sim \epsilon^{-1/2}
\, , \qquad \pp{v_2}{y_2} , \pp{v_3}{y_3} , \pp{v_2}{y_3} , \pp{v_3}{y_3} \sim O(1)\, , \qquad \epsilon \to 0\, ,
\end{split}
\label{v-bu-r1lvl1}
\end{equation}
which corresponds to a matrix $C$ given by
\begin{equation}
C=\begin{pmatrix}
 \epsilon^{-2/3}  \eta_{11} & \epsilon^{-1/2} \eta_{12} & \epsilon^{-1/2}  \eta_{13}  \\
 \epsilon^{-1/2}  \eta_{21} &   \eta_{22}  & \eta_{23} \\
 \epsilon^{-1/2}  \eta_{31} &   \eta _{32} &  \eta_{33} 
\end{pmatrix}
\end{equation}
where $\eta_{ij}$ are certain coefficients depending on $\pp{f_i}{u_j}(\uu_b)$ and $\frac{\partial^2 f_i}{\partial u_j \partial u_k}(\uu_b)$ evaluated at the point 
$\uu_b$. Consequently, the components $\omega_i$ of the vorticity have the following behavior at the blowup point of the second level
\begin{equation}
\begin{split}
\omega_1=&\epsilon^{-1/2}\tilde{S}_1(\uu_b)+\tilde{T}_1(\uu_b)\, ,  \\
\omega_2= &\epsilon^{-2/3}Y_2(\uu_b)+\epsilon^{-1/2}\tilde{S}_2(\uu_b)+\tilde{T}_2(\uu_b)\, ,\\
\omega_3= &\epsilon^{-2/3}Y_3(\uu_b)+\epsilon^{-1/2}\tilde{S}_3(\uu_b)+\tilde{T}_3(\uu_b)\, , \qquad \epsilon \to 0
\end{split}
\label{vort-bu-r1lvl1}
\end{equation}
where $Y_i,\tilde{S}_i,$ and $\tilde{T}_i$  are  certain bounded functions of $\uu_b \in \Gamma$. 
In this case the direction of vorticity vector (\ref{vort-bu-dir}) is
\begin{equation}
\boldsymbol{\hat{\omega}} = \frac{1}{|\mathbf{Y}|} \Big( 0,Y_2(\uu_b),Y_3(\uu_b) \Big)\, ,
\end{equation}
where $|\mathbf{Y}|^2=Y_2(\uu_b)^2+Y_3(\uu_b)^2 $.
\par 
So, in contrast to the first level (\ref{vort-sing-comp}) the components  of the vorticity vector generically blows up in a different manner.
Such realization occurs in the two-dimensional subspace of the blowup hypersurface $\Gamma$ \cite{KOpre}. So, one can impose at most 
two constraints.\par

Under the constraint 
\begin{equation}
\tilde{S}_1(\uu_b)=0
\label{vort-bu-011-cond}
\end{equation}
one has the following behavior 
\begin{equation}
\boldsymbol{\omega} \sim (O(1),\epsilon^{-2/3},\epsilon^{-2/3})\, , \qquad \epsilon \to 0.
\end{equation}

If instead 
\begin{equation}
Y_2(\uu_b)=0
\label{vort-bu-001-cond}
\end{equation}
then
\begin{equation}
\boldsymbol{\omega} \sim ( \epsilon^{-1/2},\epsilon^{-1/2},\epsilon^{-2/3} )\, , \qquad \epsilon \to 0.
\end{equation}
and 
\begin{equation}
\boldsymbol{\hat{\omega}} = \big( 0,0,1\big)\, . 
\end{equation}
The situations (\ref{vort-bu-011-cond}) and (\ref{vort-bu-001-cond}) may happen on curves belonging to $\Gamma_2$.\par

Imposing two constraints, one may have essentially two different situations. Indeed if 
\begin{equation}
Y_2(\uu_b)=Y_3(\uu_b)=0
\label{vort-bu-111-cond}
\end{equation}
all components of vorticity blow up in the same manner, namely,
\begin{equation}
\boldsymbol{\omega} \sim ( \epsilon^{-1/2},\epsilon^{-1/2},\epsilon^{-1/2} )\, , \qquad \epsilon \to 0.
\end{equation}
and the vorticity direction vector is generic one.
On the other hand, if it happens that
\begin{equation}
\tilde{S}_1(\uu_b)=Y_2(\uu_b)=0\, ,
\end{equation}
then the components of vorticity behave quite differently since
\begin{equation}
\boldsymbol{\omega} \sim ( O(1),\epsilon^{-1/2},\epsilon^{-2/3} )\, , \qquad \epsilon \to 0.
\end{equation}
In this case the vorticity direction vector $\boldsymbol{\hat{\omega}}$ is oriented along the third axis, namely
\begin{equation}
\boldsymbol{\hat{\omega}} = \big( 0,0,1\big)\, . 
\end{equation}
Such situation is realisable in principle at the points of intersection of the curves defined by (\ref{vort-bu-001-cond}) and  (\ref{vort-bu-011-cond}).\par

One observes similar behaviors of vorticity in other subsectors of the second level of blowups. \par

Third level of blowups is realisable  on a curve belonging to $\Gamma$. Derivatives $\pp{v_\alpha}{y_\beta}$ behave similar to (\ref{v-bu-r1lvl1}) 
except that 
\begin{equation}
\pp{v_1}{y_1} \sim \epsilon^{-3/4}\, ,
\end{equation}
and, as a consequence, one has the behavior of the type (\ref{vort-bu-r1lvl1}) with the substitution $\epsilon^{-2/3} \to \epsilon^{-3/4}$
in the $Y_i$-terms. In this case one can impose, generically, only one constraint. For instance, if $Y_3(\uu_b)=0$ one has the following
behavior of component of vorticity
\begin{equation}
\boldsymbol{\omega}=(\epsilon^{-1/2},\epsilon^{-3/4},\epsilon^{-1/2})\, , \qquad \epsilon \to 0\, . 
\end{equation}
and
\begin{equation}
\boldsymbol{\hat{\omega}} = \big( 0,1,0\big)\, .
\end{equation}
 Finally, the fourth level may occur at a point on $\Gamma$ and this point (see also \cite{KOpre}) 
 \begin{equation}
 \pp{v_1}{y_1} \sim \epsilon^{-4/5}\, , \qquad \epsilon \to 0.
 \end{equation}
Again, one has formula (\ref{vort-bu-r1lvl1}) with the substitution $\epsilon^{-2/3} \to \epsilon^{-4/5}$ in the first term in the r.h.s. and
generically no constraints are allowed.


\section{Vorticity for two-dimensional HEE}
\label{sec-vort2D}

For the two-dimensional HEE an analog of the formula (\ref{vortivel})  for the vorticity $\omega_3=\pp{u_2}{x_1}-\pp{u_1}{x_2}$ is given by
\begin{equation}
\omega_3(t,\uu)=\frac{\pp{f_1}{u_2}-\pp{f_2}{u_1}}{t^2+\tr(M_0) t +\det\left(M_0 \right)}\, ,
\label{2D-vort}
\end{equation}
where $M_0\equiv M(t=0,\uu)$ is the matrix with components $(M_0)_{ij}=\pp{f_i}{u_j}$, $i,j=1,2$.  
 The quadratic equation $t^2+\tr(M_0) t +\det\left(M_0 \right)=0$, defining the blowup surface $\Gamma$ \cite{KO22},  may have, obviously, either two 
 real roots or no one, depending on the sign of the discriminant 
 \begin{equation}
 \Delta(\uu)=\left(\pp{f_1}{u_1}+\pp{f_2}{u_2} \right)^2-4\left(\pp{f_1}{u_1} \pp{f_2}{u_2}-\pp{f_1}{u_2}\pp{f_2}{u_1} \right) = 
  \left(\pp{f_1}{u_1}-\pp{f_2}{u_2} \right)^2+4\pp{f_1}{u_2}\pp{f_2}{u_1} \, .
 \end{equation}
So, in contrast to the three-dimensional HEE, in two dimensions there are solutions  with  blowups free vorticity (cf. \cite{KO22}).\par

It is natural to consider subdomains $\mathcal{D}^+_\uu \subset \mathcal{D}_\uu$, and $\mathcal{D}^-_\uu \subset \mathcal{D}_\uu$ defined as follows
\begin{equation}
\begin{split}
(u_1,u_2) \in \mathcal{D}^+_\uu\, , \qquad \mathrm{if} \quad \Delta(u_1,u_2) >0\, , \\
(u_1,u_2) \in \mathcal{D}^-_\uu\, , \qquad \mathrm{if} \quad \Delta(u_1,u_2) <0\, , \\
(u_1,u_2) \in \mathcal{D}^0_\uu\, , \qquad \mathrm{if} \quad \Delta(u_1,u_2) =0\, , \\
\end{split}
\end{equation}
then 
\begin{equation}
\mathcal{D}_\uu= \mathcal{D}^+_\uu \cup \mathcal{D}^-_\uu \cup \mathcal{D}^0_\uu
\end{equation}
and the curve $\mathcal{D}^0$ is the boundary between $\mathcal{D}^+_\uu$ and $ \mathcal{D}^-_\uu$.
In the case $\mathcal{D}_\uu= \mathcal{D}^-_\uu$, one has the blowup free situation. \par

In the rest of this section we will assume that the subdomain $ \mathcal{D}^+_\uu$ is not empty and hence the blowup surface has two branches $
\Gamma_+$ and $\Gamma_-$.\par

Let $\uu_b$ a point at $\mathcal{D}^+_\uu$
and $t_b$ be the corresponding value of time $t$ on the first or the second branches of $\Gamma$. In the first regime, i.e. when $t \to t_b$ with fixed 
$\uu_b$, one has 
\begin{equation}
\omega_3(t_b+\varepsilon ,\uu_b)   \sim 
 \frac{\pp{f_1}{u_2}(\uu_b)-\pp{f_2}{u_1}(\uu_b)+O(\varepsilon)}
 {\left( 2t_b+  \pp{f_1}{u_1}(\uu_b)+\pp{f_2}{u_2}(\uu_b) \right) \varepsilon + \varepsilon^2  }\, , \qquad \varepsilon \to 0  \, .
\end{equation}
So, if
\begin{equation}
2t_b+  \pp{f_1}{u_1}(\uu_b)+\pp{f_2}{u_2}(\uu_b) 
=\sqrt{\Delta(\uu)}\Big{|}_{\uu=\uu_b}\neq 0
\end{equation}
the vorticity $\omega_3$ blows up as
\begin{equation}
\omega_3(t,\uu_b) \sim \varepsilon^{-1} \equiv (t-t_b)^{-1}\, , \qquad t \to t_b \, .
\label{vort2D-t-1}
\end{equation}
This happens at each point of  the blowup surface $\Gamma$. \par

If instead
\begin{equation}
2t_b+  \pp{f_1}{u_1}(\uu_b)+\pp{f_2}{u_2}(\uu_b) 
=\sqrt{\Delta(\uu)}\Big{|}_{\uu=\uu_b}= 0
\label{condvort2D-t-2}
\end{equation}
the vorticity $\omega_3$ blows up as
\begin{equation}
\omega_3(t,\uu_b) \sim \varepsilon^{-2} \equiv (t-t_b)^{-2}\, , \qquad t \to t_b \, .
\label{vort2D-t-2}
\end{equation}
Such a behavior (\ref{vort2D-t-2})  occurs the curve defined by the condition (\ref{condvort2D-t-2}). 

It is the condition of coincidence for the values  ${t_b}_\pm=\frac{1}{2} $
\begin{equation}
{t_b}_\pm=\frac{-\tr M_0 \pm \sqrt{\Delta}}{2} 
\end{equation}
of the branches $\Gamma_\pm$, i.e. ${t_b}_+={t_b}_-$. Hence, the blow-up of the type (\ref{vort2D-t-2}) occurs along the curve of intersection of two 
branches of the blow-up surface $\Gamma$. The corresponding curve (\ref{condvort2D-t-2}) in the hodograph space can be the border curve between two 
subdomains  $\mathcal{D}^+_\uu$ or $\mathcal{D}^-_\uu$ when $\mathcal{D}^+_\uu=\mathcal{D}_\uu$ or $\mathcal{D}^-_\uu=\mathcal{D}_\uu$
respectively.  \par

Similar to the three-dimensional case one can view the conditions (\ref{condvort2D-t-2}) as the equation which defines those functions $f_1(\uu)$
and $f_2(\uu)$  for which two branches of $\Gamma$ coincide. 

In order to analyze the behavior of the vorticity $\omega_3$ at fixed time $t_b$, 
similar to (\ref{bu-var}), one introduces the variables $\mathbf{y}$ and $\mathbf{v}$  (see also \cite{KOpre})
\begin{equation}
\delta \uu=\sum_{\alpha=1}^2 \mathbfcal{R}^{(\alpha)} \delta v_\alpha\, , \qquad \delta \ux=\sum_{\alpha=1}^2 \mathbfcal{P}^{(\alpha)} \delta y_\alpha\, .
\end{equation}
At the first level of blowups one has the following behavior of derivatives \cite{KOpre}
\begin{equation}
\pp{v_1}{y_1} ,
\pp{v_1}{y_2} ,
\pp{v_2}{y_1} \sim \epsilon^{-1/2}\, , \qquad 
\pp{v_2}{y_2} \sim O(1)\, .
\end{equation}
So, one has the relation 
\begin{equation}
\delta v_\alpha= \sum_{\beta=1}^2 C_{\alpha \beta} \delta y_\beta \, , \qquad \alpha=1,2,
\end{equation}
with the matrix
\begin{equation}
C=\begin{pmatrix}
\epsilon^{-1/2} \nu_{11} & \epsilon^{-1/2} \nu_{12} \\
\epsilon^{-1/2} \nu_{21}& \nu_{22}
\end{pmatrix}\, .
\end{equation}
In the two-dimensional case the vorticity is  the differential  two-form  
\begin{equation}
\omega(\uu_b)=\omega_{12} \delta y_1 \wedge \delta y_2\, ,
\end{equation}
where 
\begin{equation}
\omega_{12} (\uu_b)=\epsilon^{-1/2} S(\uu_b) + T(\uu_b)\, , \qquad \epsilon \to 0
\label{vort-bu-2D}
\end{equation}
and $S(\uu_b)$ and $T(\uu_b)$ are certain combinations of $\nu_{\alpha \beta}$ and $\mathbfcal{R}^{\alpha} \cdot \mathbfcal{P}^{\beta}$ 
(see in analogy the three-dimensional case the (\ref{vort-bu-coeff-velpos}) and (\ref{vort-bu-2f-adapcoord}) relations). \par

In the cartesian coordinates the vorticity $\pp{u_2}{x_1}-\pp{u_1}{x_2}$ also is of the form (\ref{vort-bu-2D}). Along the curve defined by the condition
\begin{equation}
S(\uu_b)=0
\end{equation}
the vorticity is bounded. \par

Blowups of second level occurs on the curve contained in $\Gamma$ and on this curve (see \cite{KOpre})
\begin{equation}
\pp{v_1}{y_1}\sim \epsilon^{-2/3}\, , \qquad
\pp{v_1}{y_2}\sim \epsilon^{-1/2}\, , \qquad
\pp{v_2}{y_1}\sim \epsilon^{-1/2}\, , \qquad
\pp{v_2}{y_2}\sim O(1)\, , \qquad \epsilon \to 0
\end{equation}
and, consequently, the vorticity blows-up as
\begin{equation}
\omega_{12}= \epsilon^{-2/3}Y(\uu_b)+\epsilon^{-1/2}S(\uu_b)+T(\uu_b)\, .
\end{equation}
Finally, at the third level which may occur at a point on $\Gamma$, one has $\pp{v_1}{y_1}\sim \epsilon^{-3/4}$ and, hence, 
 the vorticity blows-up as $\omega_{12}= \epsilon^{-3/4}$.\par
 

\section{Examples in 2D.}
\label{sec-vortexe2D}
Here we will  present three characteristic examples for the two-dimensional HEE.
\subsection{Blowup free solutions}
Let the functions $f_1$ and $f_2$ be of the form
\begin{equation}
f_1=\pp{W}{u_2}\, , \qquad f_2=\pp{W}{u_1}
\label{pot2Dexe}
\end{equation}
where the real function $W(u_1,u_2)$ obeys  the Laplace equation 
\begin{equation}
\frac{\partial^2 W}{\partial u_1^2}+
\frac{\partial^2 W}{\partial u_2^2}=0\, .
\label{potharm}
\end{equation}
It is easy to see that in this case
\begin{equation}
t_b= -\frac{\partial^2 W}{\partial u_1\partial u_2} \pm \sqrt{\Delta }   \quad \mathrm{with} \quad
\Delta =- 4 \left( \frac{\partial^2 W}{\partial u_1^2} \right)^2<0
\end{equation}
for any function $W$ except a linear one. So, the corresponding solutions $u_1$ and $u_2$ of the 2D HEE have no blowups. \par

The vorticity (\ref{2D-vort}) is given by
\begin{equation}
\omega_3 = -2 \frac{ \frac{\partial^2 W}{\partial u_1^2} }{t^2+2 \frac{\partial^2 W}{\partial u_1\partial u_2} t 
+  \left( \frac{\partial^2 W}{\partial u_1 \partial u_2} \right)^2+\left( \frac{\partial^2 W}{\partial u_1^2} \right)^2} \, .
\label{harmvort}
\end{equation}
and it is blowup free too. \par 

The particular choice 
\begin{equation}
W=\frac{1}{2\alpha} \left( u_2^2-u_1^2 \right) 
\end{equation}
or $f_1=\frac{u_2}{\alpha}$, $f_2=-\frac{u_1}{\alpha}$ corresponds to initial velocities $u_1=\alpha x_2 $  and $u_2=-\alpha x_1 $
where $\alpha$ is an arbitrary real constant. Such initial condition gives
\begin{equation}
u_1=\frac{\alpha  (\alpha   {x_1} t-{x_2})}{\alpha ^2 t^2+1}\, , \qquad u_2=\frac{\alpha  (\alpha  {x_2} t+{x_1})}{\alpha ^2 t^2+1}\, ,
\label{sol-vort-2D}
\end{equation}
and 
\begin{equation}
\omega_3=\frac{2\alpha}{\alpha^2 t^2+1}\, .
\end{equation}
It is the rotational type vortex solution of the 2D HEE with the initial strenght $2 \alpha$ and $\alpha^{-1}$ as the characteristic decaying time.\par

It is worth to note that the subclass of solutions of the 2D HEE corresponding to the choice (\ref{pot2Dexe})  has a simple description 
in terms of complex coordinates \cite{KO22}
\begin{equation}
Z=x_1+ix_2\, ,\qquad V=u_1+iv_2\, , \qquad F=f_1+i f_2\, .
\end{equation}
Indeed, in these variables the conditions (\ref{pot2Dexe}) and (\ref{potharm}) are given by
\begin{equation}
F=2i\pp{W}{V}
\end{equation}
and 
\begin{equation}
\frac{ \partial W(V,\overline{V})}{\partial V \partial \overline{V}}=0\, .
\end{equation}
Since 
\begin{equation}
W(V,\overline{V})=\mathcal{W}(V)+\overline{\mathcal{W}}(\overline{V})
\end{equation}
where $\mathcal{W}(V)$ is an arbitrary analytic function (note that (\ref{pot2Dexe}) implies that $W$ is real-valued), then 
\begin{equation}
F=2i\pp{\mathcal{W}(V)}{V}\, .
\end{equation}
For such function $F$ the hodograph equation assume the form
\begin{equation}
Z-V t = F(V)\, .
\label{hodo2Dcomplex}
\end{equation}
Solutions of the hodograph equation (\ref{hodo2Dcomplex}) obeys the equation
\begin{equation}
\pp{V}{t}+V\pp{V}{Z}=0\, .
\label{complexBH}
\end{equation}
In the complex variables the vorticity (\ref{harmvort}) is given by 
\begin{equation}
\omega_3=-2\frac{\mathrm{Im} \left( \pp{F}{V}  \right)}{ \big{\vert}t+  \pp{F}{V} \big{\vert}^2}\, .
\end{equation}
For the solution (\ref{sol-vort-2D})  $F = - {i V}/{\alpha}$.  
For the generic analytic  function $F (V)$ the corresponding  solution $V(Z,t)$ of the equation 
(\ref{complexBH}) and its vorticity  are blowup free. In the trivial particular case  $F = \beta V$, where $\beta$ is an arbitrary real constant, the solution
 $V(Z,t) = \frac{Z}{t+ \beta}$ of equation (\ref{complexBH}) and its derivative exhibit the blowup at $t = -\beta$ while the vorticity  $\omega _3 = 0$. 
 In this case the 2D HEE is decomposed into two one-dimensional Burgers-Hopf equations.\par
 
 The fact that for the generic analytic solutions of the 2D HEE the derivatives are blowups free has been noted in \cite{KO22} (Section 5). In different
 contexts the equation (\ref{complexBH}) has been considered earlier in \cite{KSZ94,KZ14,ZK18}.
\subsection{Nongeneric blowup}
\begin{figure}[ht!]
\begin{center}
\includegraphics[width=.3 \textwidth]{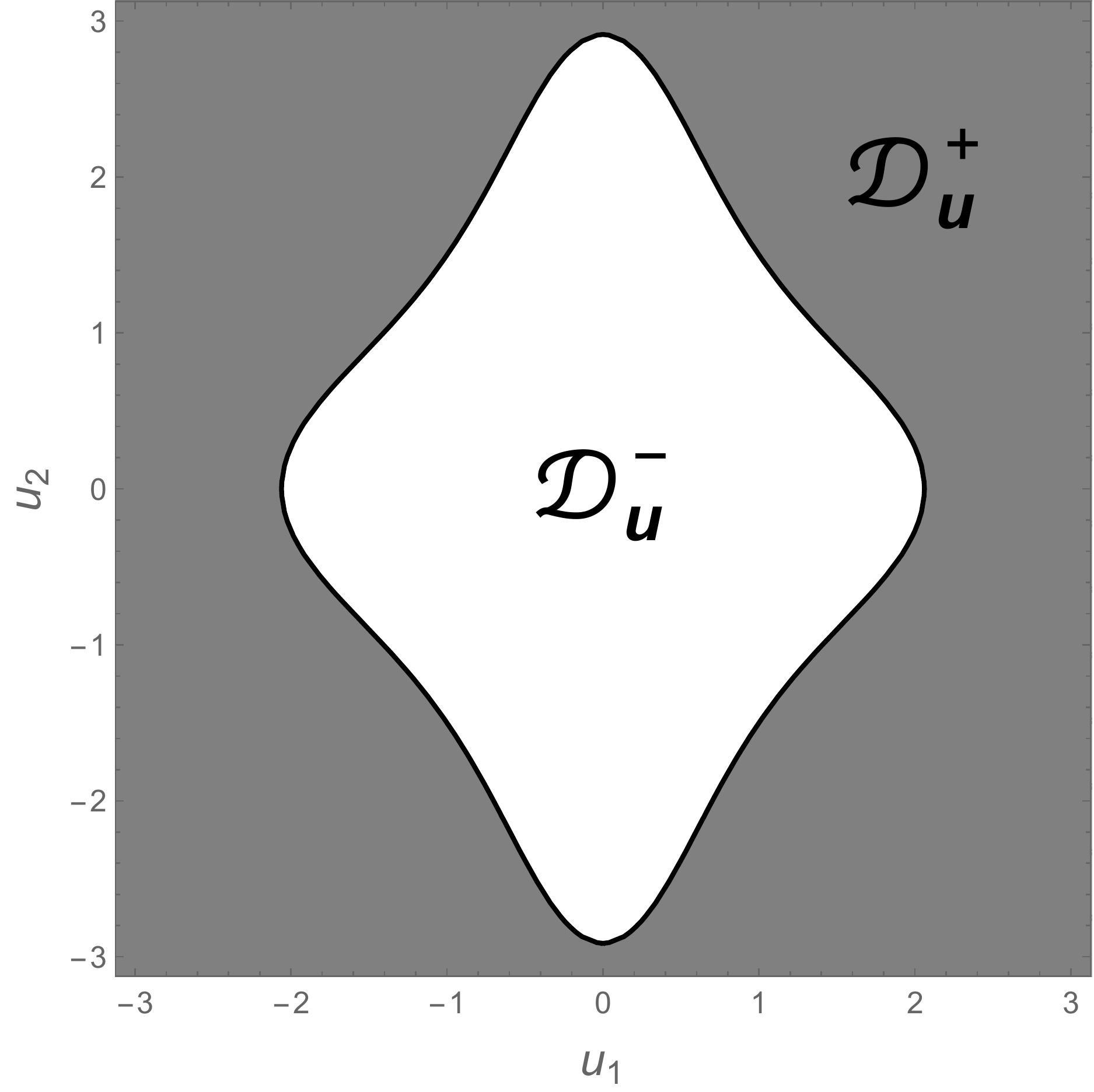}
\caption{In the gray  $\mathcal{D}_\uu^+$ region the discriminant $\Delta(u_1,u_2)$ (\ref{3map-discr}) is positive and therefore blowups are possible. 
In the complementary region $\mathcal{D}_\uu^-$ the discriminant $\Delta(u_1,u_2)$ is negative and therefore no blowups are possible.}
\label{cubic-map-domain-fig}
\end{center}
\end{figure}

\begin{figure}[ht!]
\begin{center}
\includegraphics[width=.5 \textwidth]{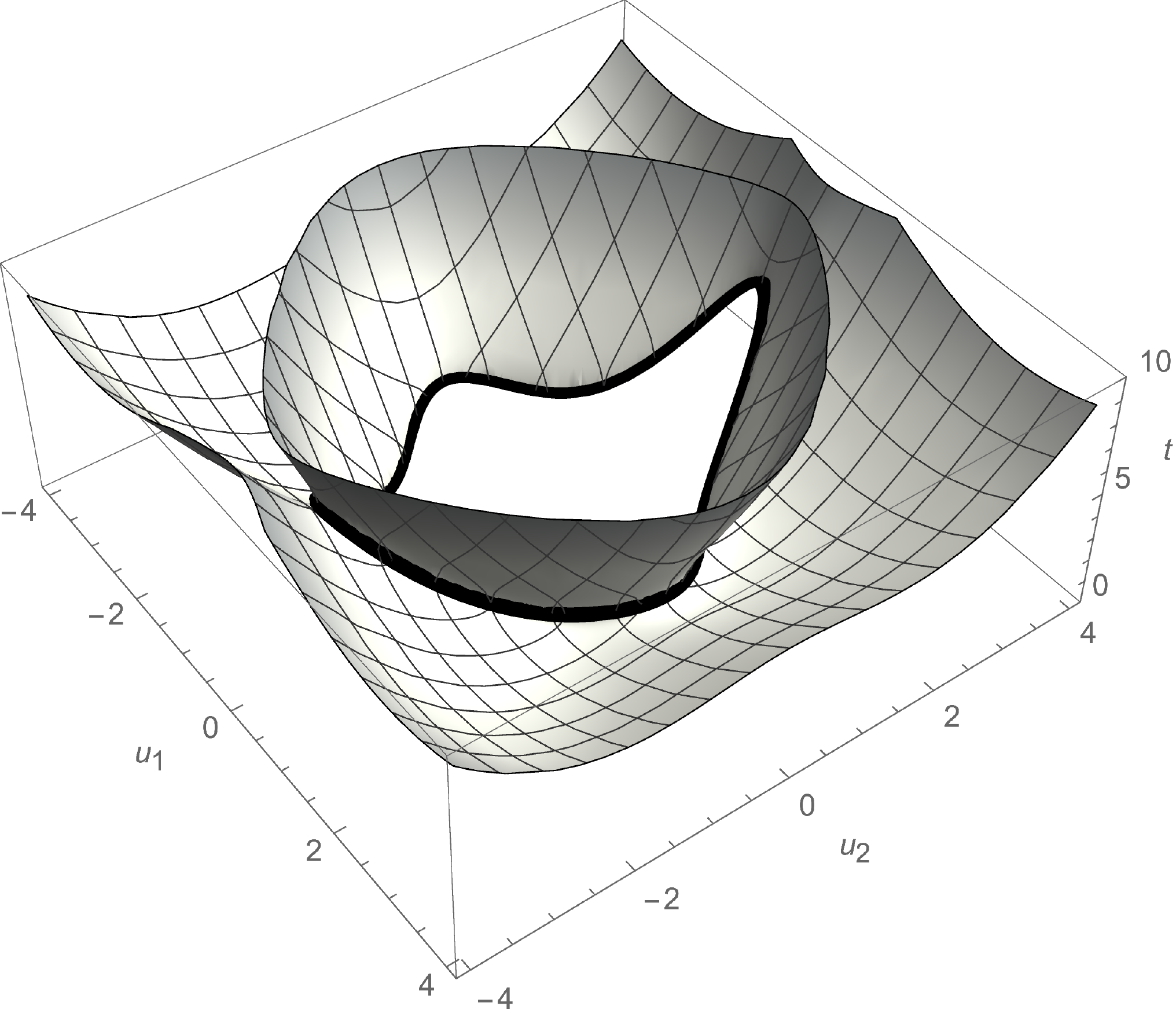}
\caption{Blowup $\Gamma$ region (\ref{Cubic-bure}) related to hodograph mappings (\ref{cubic-map-exe}). At the black curve 
(\ref{curve-nongener-vort}) the vorticity behavior is 
nongeneric $\omega \sim (\Delta t)^{-2}$}
\label{cubic-map-eigen-fig}
\end{center}
\end{figure}

Let  us choose
\begin{equation}
f_1=-\frac{u_1^3}{3}-\frac{2}{3} u_1 u_2^2 +2 u_2\, , \qquad f_2=-\frac{u_2^3}{3}-\frac{1}{3} u_1^2 u_2-u_1\, .
\label{cubic-map-exe}
\end{equation}
The corresponding initial data are 
\begin{equation}
\begin{split}
u_1(x_1,x_2,0)=& -x_2-\frac{x_1^3}{24}-\frac{1}{6} x_1 x_2^2+\frac{x_2^5}{18}+\frac{1}{72} x_1^2 x_2^3+\frac{1}{144} x_1^4 x_2 + \dots\, , \\
u_2(x_1,x_2,0)=& \frac{x_1}{2}-\frac{x_2^3}{6}-\frac{1}{12} x_2 x_1^2-\frac{x_1^5}{288}-\frac{1}{144} x_2^2 x_1^3-\frac{1}{36} x_2^4 x_1 +\dots\, .
\end{split}
\end{equation}
In this case the matrix $M$ is 
\begin{equation}
M(t,\uu)= 
\left(
\begin{array}{ccc}
 t-\left(u_1^2+\frac{2 }{3} u_2^2 \right)&& 2-\frac{4 }{3} u_1 u_2 \\ 
 && \\
 -\frac{2 }{3} u_1 u_2 - 1 && t-\left(\frac{1}{3}u_1^2+u_2^2\right) \\
\end{array}
\right)\, .
\end{equation}
and the blowup surface $\Gamma$ is defined by the equation
\begin{equation}
t^2-\left( \frac{4}{3} u_1^2+\frac{5}{3} u_2^2 \right) t +\frac{1}{3}u_1^4 + \frac{1}{3}u_1^2u_2^2 +\frac{2}{3}u_2^4+2=0\, , 
\end{equation}
The discriminant $\Delta(u_1,u_2)$ is
\begin{equation}
\Delta(u_1,u_2) \equiv 4 u_1^4+28 u_2^2 u_1^2+u_2^4-72\, .
\label{3map-discr}
\end{equation}
 So  the subdomains $\mathcal{D}^+_\uu$ and $\mathcal{D}^-_\uu$ in $\mathcal{D}_\uu$ are separated by the quartic curve
 \begin{equation}
\Delta(u_1,u_2)=4 u_1^4+28 u_2^2 u_1^2+u_2^4-72=0\, .
 \label{3map-nongeneric}
\end{equation}
The subdomain $\mathcal{D}^-_\uu$ is located around the origin $u_1=u_2=0$  as shown in figure \ref{cubic-map-domain-fig}.
The blowup surface $\Gamma$ has two branches
\begin{equation}
t_\pm= \frac{1}{6} \left(4 u_1^2+5 u_2^2\pm\sqrt{4 u_1^4+28 u_2^2 u_1^2+u_2^4-72}\right)\, .
\label{Cubic-bure}
\end{equation}
with $\uu \in \mathcal{D}_\uu^+$. It is easy to see that for both branches  $t_+\geq t_->0$ (see figure (\ref{cubic-map-eigen-fig})). 
The time of the gradient catastrophe  is ${t_-}_\mathrm{min}=1.62019$ at the point $u_1=\pm1.59562$ , $u_2=\pm1.17844$. \par

The vorticity is equal to
\begin{equation}
\omega_3=\frac{3-\frac{2}{3}u_1u_2}{\det M}\, .
\end{equation}

In the first regime of approaching of generic  blowup point $(u_1,u_2) \in \mathcal{D}^+_\uu$ the vorticity behaves as  
\begin{equation}
\omega_3(t,\uu_b) \sim \pm \frac{2 u_1 u_2-9}{\sqrt{4 u_1^4+28 u_2^2 u_1^2+u_2^4-72}
   \left(t-t_b\right)}\, , \qquad t \to t_b= \frac{1}{6} \left(4 u_1^2-5 u_2^2\pm\sqrt{4 u_1^4+28 u_2^2
   u_1^2+u_2^4-72}\right)\, .
\end{equation}
Approaching the points 
\begin{equation}
{t_\pm}_b=\frac{2}{3}u_1^2+\frac{5}{6}u_2^2\, , \qquad \Delta(u_1,u_2)=0\, ,
\label{curve-nongener-vort}
\end{equation}
which belongs to the curve of intersection of two branches $t_+$ and $t_-$, the vorticity blows up as
\begin{equation}
\omega \sim \, \frac{\pm\frac{2}{3} \sqrt{2} u_1 \sqrt{-7 u_1^2-\sqrt{6} \sqrt{8 u_1^4+3}}-3}{\left(t-t_b\right)^2}\, , \qquad
t \to t_b=-11 u_1^2-5 \sqrt{\frac{16 u_1^4}{3}+2} \,.
\end{equation}
In this case the curve $ \Delta(u_1,u_2)=0$ is the boundary line between the subdomains $\mathcal{D}_\uu^+$ and $\mathcal{D}_\uu^-$.
\subsection{Gaussian initial data}
 Finally we consider solution of the HEE with the  initial data
\begin{equation}
u_1(\ux,0)=e^{-x_1^2-x_2^2}\, , \qquad  u_2(\ux,0)=\exp^{-x_1^2-3x_2^2}\, . 
\label{inival-exe}
\end{equation}
Such initial values admits four different open sets of invertibility shown in the  table \ref{inidatagauss-2D-exe}.
\begin{table}
\begin{equation*}
\begin{array}{|c|c|c|c|c|}
\hline
& \begin{array}{c}{x_1 \geq 0}  \vspace{.1cm}  \\ { x_2 \geq 0} \end{array}&  \begin{array}{c}{x_1 \geq 0}  \vspace{.1cm} \\{ x_2 < 0} \end{array} & 
\begin{array}{c}{x_1 < 0}  \vspace{.1cm}  \\{ x_2 \geq 0} \end{array}  & \begin{array}{c}{x_1 < 0}  \vspace{.1cm}  \\{ x_2 < 0} \end{array} \\

\hline
&&&&\\
x_1=f_1(\uu) & 
\sqrt{\frac{1}{2} \log \frac{u_2}{{u_1}^3}} &\sqrt{\frac{1}{2} \log \frac{u_2}{{u_1}^3}}&-\sqrt{\frac{1}{2} \log \frac{u_2}{{u_1}^3}}&-\sqrt{\frac{1}{2} \log \frac{u_2}{{u_1}^3}} \\
&&&&\\
\hline
&&&&\\
x_2=f_2(\uu) &
\sqrt{\frac{1}{2} \log \frac{u_1}{{u_2}}}&-\sqrt{\frac{1}{2} \log \frac{u_1}{{u_2}}}&\sqrt{\frac{1}{2} \log \frac{u_1}{{u_2}}}&-\sqrt{\frac{1}{2} \log \frac{u_1}{{u_2}}} \\
&&&&\\
\hline
\end{array}
\end{equation*}
\caption{The local inverses of the initial data (\ref{inival-exe}).}
\label{inidatagauss-2D-exe}
\end{table}
where $f_i$, $i=1,2$ is the local inverse of (\ref{inival-exe}).
The hodograph equations (\ref{hodogen}) assume the form of the system of four equations
\begin{equation}
G_{a,b}: \left\{  \begin{array}{cc} 
							 x_1=u_1t +a\sqrt{\frac{1}{2} \ln \left( \frac{u_2}{u_1^3}\right)}\, , & a (x_1-u_1t)>0  \\
							  x_2=u_2t +b  \sqrt{\frac{1}{2} \ln \left( \frac{u_2}{u_1}\right)}\, , & b (x_2-u_2t)>0  \\
			    \end{array}   \right. \, ,  \qquad a=\pm\,  , b=\pm\, .
\label{hodo-exp-exe}			    
\end{equation}
Each pair of equations (\ref{hodo-exp-exe}) define a solution $\uu_{ab}(\ux,t)$ 
in the corresponding subdomain. So, solution of the 2D HEE with the initial 
data (\ref{inival-exe}) is a union 
\begin{equation}
\uu(\ux,t)= \uu_{++}(\ux,t) \cup \uu_{+-}(\ux,t) \cup \uu_{-+}(\ux,t) \cup \uu_{--}(\ux,t) \, .
\end{equation}
In other words
\begin{equation}
\uu(\ux,t)=
\left\{ 
\begin{array}{c}
 \uu_{++}(\ux,t) \, , \qquad  \mathrm{at} \quad x_1-u_1(\ux,t)>0\, , \quad x_2-u_2(\ux,t)>0\, ,\\
 \uu_{+-}(\ux,t)  \, , \qquad  \mathrm{at} \quad x_1-u_1(\ux,t)>0\, ,\quad x_2-u_2(\ux,t)<0\, ,\\
  \uu_{-+}(\ux,t) \, , \qquad  \mathrm{at} \quad  x_1-u_1(\ux,t)<0\, ,\quad x_2-u_2(\ux,t)>0\, ,\\
   \uu_{--}(\ux,t)  \, , \qquad   \mathrm{at} \quad x_1-u_1(\ux,t)<0\, ,\quad x_2-u_2(\ux,t)<0\, .
\end{array}
\right.
\label{soltotexeexp}
\end{equation}
The function (\ref{soltotexeexp}) is continuous on $\mathbb{R}^2 \times \mathbb{R} $ through the boundary $\ux-\uu t=0$.
Note that $\uu_{--}(\ux,t)=\uu_{++}(-\ux,-t)$, $\uu_{-+}(\ux,t)=\uu_{+-}(-\ux,-t)$. Moreover the domain $\mathcal{D}_\uu$ is the square 
$0< u_1(\ux,t),u_2(\ux,t) \leq 1$. 
Using the standard formulae $\uu(x,t)=\uu_0(\xi_1,\xi_2)$ with $\xi_i=x_i-u_it$, $i=1,2$, one can view the piecewise solution (\ref{soltotexeexp})
as
\begin{equation}
\begin{split}
\uu_{ab}(x,t)=\uu_0(a\xi_1,b\xi_2)\, , \qquad  a\xi_1>0\, , \quad b\xi_2>0 \, .
\end{split}
\end{equation}

 Then four corresponding matrices $M$ are of the form 
\begin{equation}
M^{(ab)}(t,\uu)=\left(
\begin{array}{ccc}
 t-  a\frac{3}{2 \sqrt{2} {u_1} \sqrt{\log \left(\frac{{u_2}}{{u_1}^3}\right)}} &&
   a\frac{1}{2 \sqrt{2} {u_2} \sqrt{\log \left(\frac{{u_2}}{{u_1}^3}\right)}} \\&&\\
 b\frac{1}{2 \sqrt{2} {u_1} \sqrt{\log \left(\frac{{u_1}}{{u_2}}\right)}} &&
   t-b\frac{1}{2 \sqrt{2} {u_2} \sqrt{\log \left(\frac{{u_1}}{{u_2}}\right)}} \\
\end{array}
\right)\, , \qquad a,b=\pm
\end{equation}
and the corresponding branches of the blowup surface are defined by the equation 
\begin{equation}
\det M = t^2
-\left( \frac{3a}{2 \sqrt{2} {u_1} \sqrt{\log \left(\frac{{u_2}}{{u_1}^3}\right)}} +
\frac{b}{2 \sqrt{2} {u_2} \sqrt{\log \left(\frac{{u_1}}{{u_2}}\right)}}
\right) t 
+ \frac{ab}{4 {u_1u_2} \sqrt{\log \left(\frac{{u_2}}{{u_1}^3}\right)\log \left(\frac{{u_1}}{{u_2}}\right)}} =0\, , \quad a,b=\pm\, .
\label{bu-region-exp-exe}
\end{equation}
The values of the vorticity $\omega_3$ for the branches $(a,b)$ are given by
\begin{equation}
\omega_3= \frac{1}{\det M_{pq}(t,\uu)} \left(\frac{a}{2 \sqrt{2} {u_2} \sqrt{\log \left(\frac{{u_1}}{{u_2}^3}\right)}}- 
 \frac{b}{2 \sqrt{2} {u_1} \sqrt{\log \left(\frac{{u_2}}{{u_1}}\right)}} \right) \, .
 \label{vortivel-exp-exe}
\end{equation}
The discriminant $\Delta$ of the equation (\ref{bu-region-exp-exe}) is positive for all values of $a$ and $b$ since
\begin{equation}
\begin{split}
\Delta_{ab}(\uu)=& \left( \frac{3a}{2 \sqrt{2} {u_1} \sqrt{\log \left(\frac{{u_2}}{{u_1}^3}\right)}} +
\frac{b}{2 \sqrt{2} {u_2} \sqrt{\log \left(\frac{{u_1}}{{u_2}}\right)}}
\right)^2- \frac{ab}{ {u_1u_2} \sqrt{\log \left(\frac{{u_2}}{{u_1}^3}\right)\log \left(\frac{{u_1}}{{u_2}}\right)}} \\
=&\left( \frac{3a}{2 \sqrt{2} {u_1} \sqrt{\log \left(\frac{{u_2}}{{u_1}^3}\right)}} -
\frac{b}{2 \sqrt{2} {u_2} \sqrt{\log \left(\frac{{u_1}}{{u_2}}\right)}}
\right)^2+ \frac{ab}{ {2u_1u_2} \sqrt{\log \left(\frac{{u_2}}{{u_1}^3}\right)\log \left(\frac{{u_1}}{{u_2}}\right)}}\, .
\end{split}
\end{equation}
So, for the solution (\ref{soltotexeexp}) the blowup surface $\Gamma$ has two brances for all values of $\uu \in \mathcal{D}_\uu$
\begin{equation}
(t_{\pm})_{ab}= \frac{3a}{2 \sqrt{2} {u_1} \sqrt{\log \left(\frac{{u_2}}{{u_1}^3}\right)}} +
\frac{b}{2 \sqrt{2} {u_2} \sqrt{\log \left(\frac{{u_1}}{{u_2}}\right)}} \pm \sqrt{\Delta_{ab}}\, .
\label{tbu-branches-exp-exe}
\end{equation}
 It is easy to see that for the $(a,b)=(+,+)$ piece
 \begin{equation}
 (t_\pm)_{++}>0\, , \qquad  (t_+)_{++}> (t_-)_{++}\, ,
 \end{equation}
while
 \begin{equation}
 (t_\pm)_{--}<0\, , \qquad  (t_+)_{--}> (t_-)_{--}\, ,
 \end{equation}
For the pieces $(a,b)=(+,-)$ and $(a,b)=(-,+)$ one has
\begin{equation}
(t_+)_{+-}>0\, , \qquad (t_-)_{+-}<0\, ,
\end{equation}
and 
\begin{equation}
(t_+)_{-+}>0\, , \qquad (t_-)_{-+}<0\, .
\end{equation}
Minimal values of ${t_\pm}_{ab}$ for the positive pieces are
\begin{equation}
(t_-)_{++}\big{\vert}_{\min}= 0.642593\, , \qquad
(t_+)_{+-}\big{\vert}_{\min}= 1.16582\, , \qquad
(t_+)_{-+}\big{\vert}_{\min}= 0.673088 \, .
\end{equation}
Thus, the gradient catastrophe occurs at 
\begin{equation}
t_c \equiv (t_-)_{++}\big{\vert}_{\min}= 0.642593\, , \qquad \uu_c=(0.803494, 0.584021)\, , \qquad \ux_c=(0.759774, 0.77468)\, .
\label{cata-val-exe}
\end{equation}
\par

\begin{figure}[ht!]
\begin{center}
\includegraphics[width=.9 \textwidth]{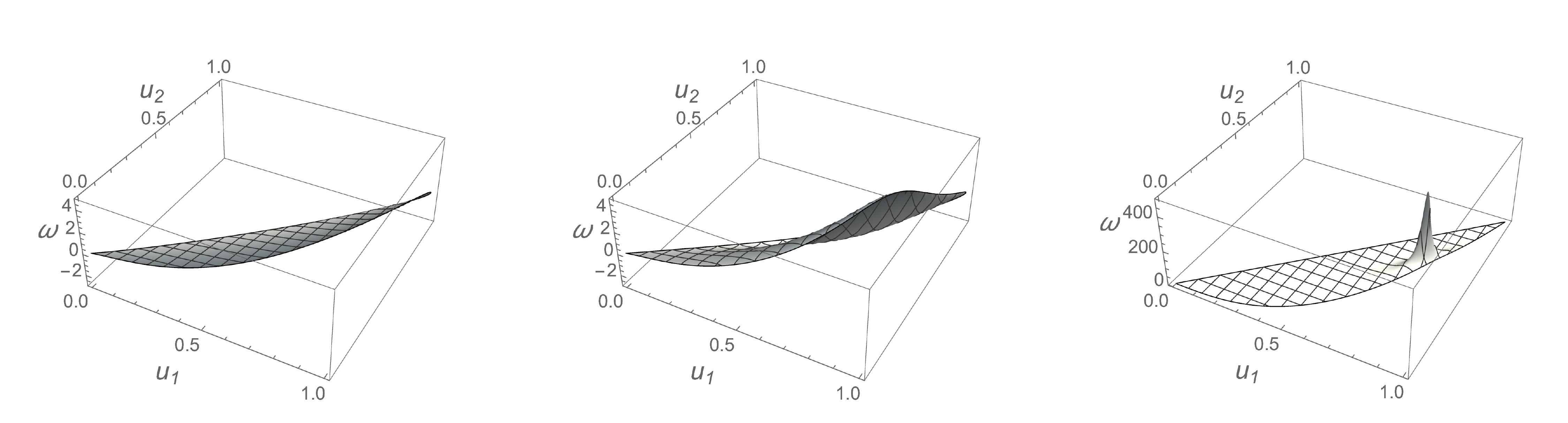}
\caption{The time evolution of the vorticity depending on $\uu$ with initial data given by (\ref{inival-exe}). From left to right the times are $t=0$, 
$t=0.85 t_c$, $t=0.999 t_c$ where $t_c=0.642593$ is the catastrophe time. Remark the change in the vertical scale in the last plot.}
\label{vortivel-exp-exe-fig}
\end{center}
\end{figure}

\begin{figure}[ht!]
\begin{center}
\includegraphics[width=.9 \textwidth]{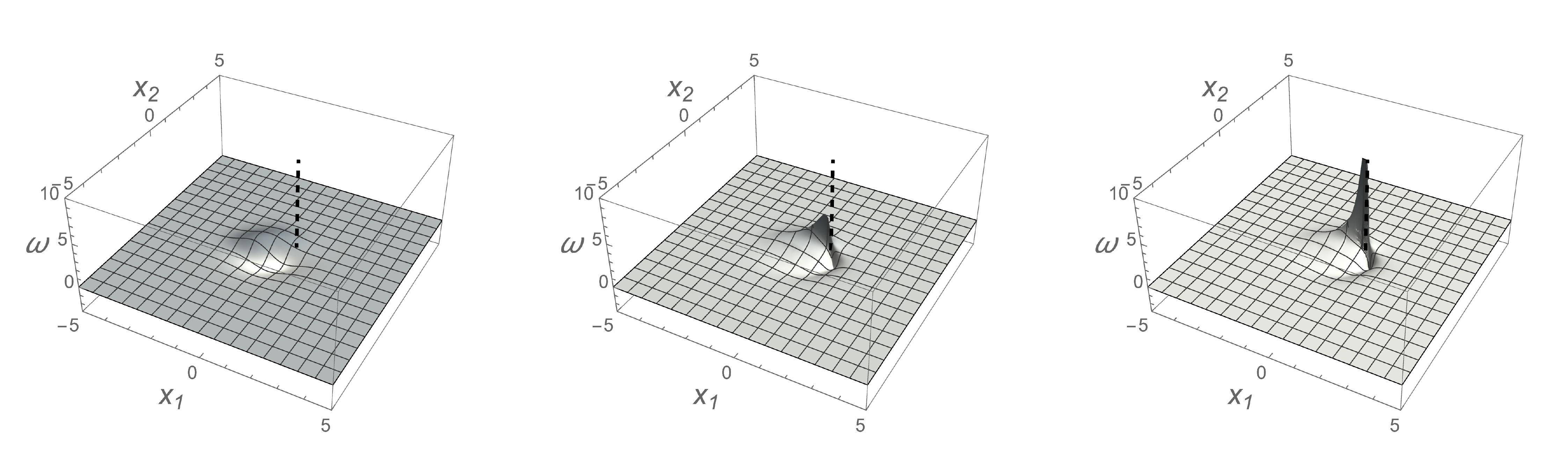}
\caption{The time evolution of the vorticity depending on $\ux$ with initial data given by (\ref{inival-exe}). From left to right the times are $t=0$, 
$t=0.85 t_c$, $t=0.999 t_c$ where $t_c=0.642593$ is the catastrophe time. The dashed vertical line indicates the catastrophe direction of the vorticity
in the catastrophe place $\ux_c$.}
\label{vortispace-exp-exe-fig}
\end{center}
\end{figure}

As expected the behavior of the vorticity at $\uu=\uu_c$ in the first regime is
\begin{equation}
\omega(t,\uu_c)=\frac{0.270466}{t_c-t} - 0.0747002 + 0.0206315 (t_c-t)+\dots\, .
\end{equation}
The time evolution of the vorticity $\omega(t,\uu)$ is shown in figure \ref{vortivel-exp-exe-fig}. 
In figure \ref{vortispace-exp-exe-fig} it is shown the time evolution of the vorticity w.r.t. to space variables, numerically computed using Mathematica.
The behavior is in agreement with the analytical predictions (\ref{cata-val-exe}).
Since $\Delta_{ab}(\uu)\neq 0$ for all $\uu \in \mathcal{D}_\uu$, two branches (\ref{tbu-branches-exp-exe}) do not intersect.
So, the blowup of the type $\omega \sim (t_c-t)^{-2}$ is absent in this case. 

\section{Blowups for n-dimensional case.}
\label{sec-n-dim}

An extension of the results presented in this paper to the $n$-dimensional HEE is quite straightforward. Indeed, the components $\omega_{ij}$
of the vorticity two-form (\ref{vort-2form}) in Cartesian coordinates are given by
\begin{equation}
\omega_{ij}(t,\uu)=(M^{-1})_{ji}(t,\uu)-(M^{-1})_{ij}(t,\uu) = \frac{\widetilde{M}_{ji}(t,\uu)-\widetilde{M}_{ij}(t,\uu)}{\det(M(t,\uu))}\, , \qquad i,j=1,\dots,n.
\end{equation}

In the $n$-dimensional case $\det(M(t,\uu))$ is a  polynomial in $t$ of degree $n$ \cite{KO22}, i.e.
\begin{equation}
\det(M(t,\uu))= \prod_{k=1}^n (t-{t_b}_k)\, .
\end{equation}
So, in the first regime of approaching the blowup point $\omega_{ij}$ may have the following behavior 
\begin{equation}
\omega_{ij} \sim (t-t_b)^{-m}\, , \qquad t \to t_b\, , \quad m=1,\dots,n\, .
\label{vort-nD-velfix}
\end{equation}
As far as the second regime is concerned it was shown in \cite{KOpre} that the derivatives $\partial u_i / \partial x_j$
may have singularities of the type $|\delta \ux|^{-\frac{m}{m+1}}$, with $m=1, \dots, n+1$. Hence, in this regime the vorticity two-form may blow up as
\begin{equation}
\omega_{ij} \sim |\delta \ux|^{-\frac{m}{m+1}}\, , \qquad |\delta \ux| \to 0\, , \qquad m=1, \dots, n+1\, .
\label{vort-nD-timefix}
\end{equation}
Similar to the results described in \cite{KOpre} blowups of the vorticity exhibit rather rich fine structure. 

The formulae (\ref{vort-nD-velfix}) and (\ref{vort-nD-timefix}) imply certain behavior of the characteristics of vorticity in different dimensions  
discussed in \cite{AK}. \par

One obtains analogous results for the stress tensor 
\begin{equation}
\mathcal{S}_{ij} \equiv \pp{u_i}{x_j}+\pp{u_j}{x_i}=(M^{-1})_{ij}+(M^{-1})_{ji} = \frac{\widetilde{M}_{ij}+\widetilde{M}_{ji}}{\det(M)}\, , \qquad i,j=1,\dots,n
\end{equation}
which is another important quantity in the theory of continuous media \cite{Lamb,L-VI,Bac}.

\section{Conclusions}
\label{sec-Con}
The results presented in this note are in part the consequences of those obtained in the paper \cite{KOpre}. As in \cite{KOpre} we are dealing with the 
most simplified version of the Navier-Stokes equation, namely with HEE (\ref{HEeq}) and do not discuss the possibility of blowups of vorticity of type 
(\ref{vort-bu}) for positive values of time. \par

All that indicates at least two possible direction of further study. The first is the verification of the realisability of hierarchy of blowups (\ref{vort-bu})  
for positive times that is of most interest in physical applications. \par

An extension of such type of analysis for more physical systems would be the second direction. In particular, it may be applicable to those 
hydrodynamical systems which are obtainable as the constraints of the multidimensional homogeneous Euler equation \cite{KO21}. \par

 

\subsubsection*{Acknowledgments}
This project has received funding from the European Union's Horizon 2020 research and innovation programme under the Marie Sk{\l}odowska-Curie grant no 778010 IPaDEGAN. We also gratefully acknowledge the auspices of the GNFM Section of INdAM, under which part of this work was carried out, and the financial support of the project MMNLP (Mathematical Methods in Non Linear Physics) of the INFN.



\end{document}